\documentclass[journal=jpca,manuscript=article]{achemso}

\usepackage{ragged2e}
\usepackage{graphicx}
\usepackage{color}
\usepackage{hyperref}
\usepackage{float}
\usepackage{array, multirow}
\usepackage{amsmath, amssymb}
\usepackage{newtxtext}              
\usepackage{booktabs}
\usepackage{xcolor} 
\usepackage{mathtools}
\usepackage{soul}
\usepackage{url}
\usepackage{qcircuit}
\usepackage{ulem}
\usepackage[T1]{fontenc}

\hypersetup{colorlinks, 
    linkcolor={blue!75!black!80!yellow},
    citecolor={blue!75!black!80!yellow}, 
    urlcolor={blue!75!black!80!yellow}
    }

\title{A perturbative non-Markovian treatment to low-temperature spin decoherence}

\author{Timothy J. Krogmeier}
\affiliation{Department of Chemistry, University of Minnesota, Minneapolis, MN 55455 USA}
\author{Anthony W. Schlimgen}
\affiliation{Department of Chemistry, University of Minnesota, Minneapolis, MN 55455 USA}
\author{Kade Head-Marsden}
\affiliation{Department of Chemistry, University of Minnesota, Minneapolis, MN 55455 USA}
\email{khm@umn.edu}

\begin{document}

\begin{abstract}
Molecular spins are promising candidates for quantum information science, leveraging coherent electronic spin states for quantum sensing and computation. However, the practical application of these systems is hindered by electronic spin decoherence, driven by interactions with nuclear spins in the molecule and the surrounding environment at low temperatures. Predicting dephasing dynamics remains a formidable challenge due to the complexity of the spin bath. In this work, we develop a non-Markovian time-convolutionless master equation to treat an electronic spin coupled to a nuclear-spin bath. By relating \textit{ab initio} electronic structure parameters directly to the decoherence dynamics, we provide a framework that accounts for pure dephasing in the low-temperature limit. We apply this method to a series of molecular qubit candidates and demonstrate good agreement with experimental relaxation trends. This approach offers a computationally efficient path for the prediction of low-temperature decoherence trends in molecular spin systems.
\end{abstract}

\maketitle


Modern technologies in quantum information science (QIS) such as quantum computing and sensing aim to leverage properties of quantum spin systems for information storage and processing beyond classical capabilities. There is significant research interest surrounding magnetically-active molecules because of their potential for room-temperature operation and the breadth of synthetic techniques available for molecular qubit design~\cite{Gatteschi:2006a,Mullin:2023a,Sanvito:2011a,Wasielewski:2020a,Coronado:2020}. Implementation of molecular spin systems typically uses an unpaired electron spin state as the qubit in QIS or perturbations of the spin states for fine magnetic field sensing.~\cite{Bayliss:2020,Gaita:2019,Yu:2021} In any application, control and manipulation of coherent spin states are necessary for robust information processing, yet present a major challenge to the viability of many quantum technologies. In low temperature molecular spin systems, fluctuations in the magnetic field interacting with the electron spin present a significant source of pure dephasing noise. Spin-active nuclei on a molecule or in a surrounding solvent system contribute a time-dependent magnetic field that can dephase the electron spin quickly.~\cite{Sproules:2016,Liu:2019,Patel:2025,Alfieri:2023} Dynamical decoupling in the form of the Hahn-echo experiment is often employed in order to remove contributions from static inhomogeneities, isolate specific decoherence mechanisms, and elongate spin dephasing time.~\cite{Hahn:1952,Lange:2010,Jiangfeng:2009} \textcolor{black}{Connecting the rates of spin dephasing to microscopic parameters requires treatment of the molecular electronic structure and the dipolar coupling between nearby nuclear spins that contribute to the Hahn-echo decay.}~\cite{Cywinski:2010,Sousa:2009,Witzel:2012} 

Current computational methods to treat electron spin dephasing include cluster-correlation expansion (CCE) approaches,~\cite{Ren-Bao:2008a, Ren-Bao:2009a, Ren-bao:2012a, Yang:2020, Chen:2020, Onizhuk:2021, Onizhuk:2024a} tensor-network methods,~\cite{Li:2025} and the analytical pair-product approximation (APPA).~\cite{Jeschke:2023,Suchaneck:2025} \textcolor{black}{Despite the breadth of different methods, there is a lack of open quantum systems (OQS) master equation (ME) treatments which explicitly relate molecular properties and dipolar coupling within the nuclear spin bath to pure dephasing dynamics.} Longitudinal relaxation, $T_1$, of the electron spin has been treated using microscopic Markovian MEs,~\cite{Albino:2019,Lunghi:2022,Lunghi:2023,Aruachan:2023} and the analogous relationship for $T_2$ dephasing has also been explored through phenomenological approaches.~\cite{Krogmeier:2024, Onizhuk:2024a} Notably microscopic MEs that explicitly relate molecular electronic structure properties to pure dephasing are desirable. \textcolor{black}{Previous studies using microscopic MEs have been applied to the central spin model, but have not incorporated the dipolar coupling between environmental nuclear spins or the effect of an applied pulse such as the Hahn-echo pulse sequence.~\cite{Ferraro:2008,Fischer:2007,Barnes:2011,Barnes:2012}} Here we derive a non-Markovian master equation for the electron spin degrees of freedom to relate \textit{ab initio} electronic structure to low-temperature decoherence trends where dephasing dominates. Specifically, we derive a time-convolutionless master equation to 2nd order perturbation theory (TCL2) with respect to the interaction Hamiltonian. \textcolor{black}{The TCL approach is a non-Markovian treatment of loss in an open system, suitable for the description of pure dephasing of an electron interacting with nearby nuclear spins.} We include a single pulse in our differential equation to better compare to Hahn-echo experiments. Through the TCL2 equation, we relate the hyperfine couplings computed with electronic structure and nuclear dipolar couplings to the coherence in the electron reduced density matrix. We obtain a result that treats the electron spin dephasing mechanism in an OQS framework with electronic structure and find that we correctly predict dephasing trends in a series of molecular qubit candidates. \textcolor{black}{The treatment of the dipolar-induced Hahn-echo decay in an OQS framework is an important step to generalize theories of spin decoherence beyond the pure dephasing limit. This could be integrated into current ME frameworks which treat $T_1$ relaxation induced by spin-phonon coupling,~\cite{Albino:2019,Lunghi:2022,Lunghi:2023,Aruachan:2023} relating \textit{ab initio} molecular properties to decoherence trends across various parameter regimes. In cases where $T_1$ and $T_2$ occur on similar timescales, it may be necessary to incorporate both mechanisms in a unified model.~\cite{McNamara:2023,Graham:2017}}

We consider an electron spin doublet interacting with two spin-$\frac{1}{2}$ nuclei. We start with the spin Hamiltonian, invoke the secular approximation, and assume a large magnetic field limit such that we are in the pure dephasing regime of spin decoherence. The electron spin Hamiltonian is,
\begin{equation}
    \hat{H}_e = \omega_e\hat{S}^z
    \label{Eq:He}
\end{equation}
where $\omega_e$ is the Larmor frequency and $\hat{S}^z$ is the Pauli-$z$ operator. The nuclear spin Hamiltonian for spins 1 and 2 is given by,
\begin{align}
    \notag
    \hat{H}_n = (\omega_1 + \frac{1}{2}A_1)\hat{I}_1^z + (\omega_2 + \frac{1}{2}A_2)\hat{I}_2^z + b_{12}\big( \hat{I}_1^z\hat{I}_2^z-\frac{1}{4}(\hat{I}_1^+\hat{I}_2^-+\hat{I}_1^-\hat{I}_2^+) \big)
    \label{Eq:Hn}
\end{align}
where $\omega_{1,2}$ is the Larmor frequency, $\hat{I}_{1,2}$ are the nuclear spin operators, $A_{1}$ and $A_{2}$ are the $zz$ components of the hyperfine tensors, and $b_{12}$ is the nuclear spin-spin dipolar coupling constant. 

The interaction between the electron spin and two nuclear spins is given by,
\begin{equation}
    \hat{H}_{en}(t) = \hat{S}^z(t)\otimes\sum_kA_k\hat{I}_k^z(t),
    \label{Eq:Hen}
\end{equation}
where $\hat{I}_k^z(t)$ evolves in time in the interaction picture according to,
\begin{equation}
    \hat{I}_k^z(t) = e^{i\hat{H}_nt}\hat{I}_k^ze^{-i\hat{H}_nt},
    \label{Eq:IP_nuclearspin}
\end{equation}
and $\hat{S}^z(t)$,
\begin{equation}
    \hat{S}^z(t) = h(t,t_p)\cdot e^{i\hat{H}_et}\hat{S}^ze^{-i\hat{H}_et}
    \label{Eq:IP_electronspin}
\end{equation}
where we have included the Heaviside step function modified to range in value from $1$ to $-1$, $h(t,t_p)=(-2\Theta(t-t_p)+1)$, to incorporate the effects of an applied pulse at time $t_p$. 

The TCL is formally derived as a series expansion of the interaction Hamiltonian, but is often taken to only a few orders. In the limit that the magnetic field is strong, the secular spin Hamiltonian conserves the total $z$-component of the spin-angular momentum. We choose a \textcolor{black}{thermal} initial state\textcolor{black}{, suitable for descriptions of pure dephasing due to nuclear spins at greater than millikelvin temperatures. Additionally, }all odd-numbered terms in the series expansion vanish, \textcolor{black}{thus} the lowest-order term required is the second-order term,
\begin{equation}
    \frac{d}{dt}\mathcal{P}\hat{\rho}(t) = \mathcal{K}_2(t)\mathcal{P}\hat{\rho}(t),
    \label{Eq:TCL_2nd_Order}
\end{equation}
where $\hat{\rho}(t)$ is the total density matrix of the composite system and environment, $\mathcal{P}$ is a projection operator that projects the density matrix onto the states of interest, and $\mathcal{K}_2(t)$ is the 2nd order TCL generator,
\begin{equation}
    \mathcal{K}_2(t) = \int_0^tdt_1\mathcal{P}\mathcal{L}(t)\mathcal{L}(t_1)\mathcal{P},
    \label{Eq:2nd_order_Kernel}
\end{equation}
where $\mathcal{L}(t)$ is the Liouvillian, $\mathcal{L}(t) = -i[\hat{H}_I(t),\hat{\rho}(t)]$. The integral in Eq.~\ref{Eq:2nd_order_Kernel} deconvolutes the density matrix in time to yield a time-local ME for $\hat{\rho}(t)$~\cite{Breuer:2007}. \textcolor{black}{Here we select the projection operator to project onto a product state of the system and the thermal state of the nuclear spins, ${\mathcal{P}\hat{\rho}=\hat{\rho}_e\otimes\rho_B}$}. \textcolor{black}{Correlated projection operators can be employed to describe entanglement between the central spin and bath spins;~\cite{Ferraro:2008,Fischer:2007} however, because the total Hamiltonian conserves the spin angular momentum, the use of the correlated projection operator is equivalent to the product state projection operators.} These considerations simplify the TCL equation,  
\begin{align}
    \frac{d\hat{\rho}\textcolor{black}{_e}(t)}{dt} = &\bigg(2\hat{S}^z\hat{\rho}\textcolor{black}{_e}\hat{S}^z-\frac{\hat{\rho}\textcolor{black}{_e}}{2}\bigg)
    \int_0^tdt_1 h(t_1,\frac{t}{2})\langle A_1\hat{I}_1^z(t)A_2\hat{I}_2^z(t_1)\textcolor{black}{\hat{\rho}_B}\rangle.
\end{align}
The off-diagonal element of the electron reduced density matrix, or the coherence, can be found analytically, 
\begin{equation}
    \rho_e^{01}(t) = \rho_e^{01}(0)e^{-W(t)},
    \label{Eq:rho01}
\end{equation}
where 
\begin{equation}
    W(t) = \int_0^tdt\int_0^tdt_1h(t_1,\frac{t}{2})\langle A_1\hat{I}_1^z(t)A_2\hat{I}_2^z(t_1)\rangle.
    \label{Eq:Wtintegral}
\end{equation}
We integrate to yield,
\begin{equation}
    W_{12}(t) = \bigg(\frac{2\Delta_{12}b_{12}}{\Delta_{12}^2+b_{12}^2}\bigg)^2 \textcolor{black}{ \textrm{sin}^4\bigg( \frac{t}{4}\sqrt{\Delta_{12}^2+b_{12}^2} \bigg)},
    \label{Eq:Wtsolved}
\end{equation}
where $\Delta_{12}=A_1-A_2$. This solution is a 2$^{nd}$ order perturbative approach to pure dephasing of an electron interacting with two nuclear spins indexed by $1$ and $2$ in a Hahn-echo experiment \textcolor{black}{with a single applied pulse occuring at $\frac{t}{2}$}. Further details of this process are shown in the Supporting Information (SI). 

To extend this approach to larger systems while retaining the direct connection to molecular parameters, we split the nuclear spin correlation function into pairs.~\cite{Sousa:2009} This yields a factorized solution for many nuclear spins,
\begin{equation}
    \rho_e^{01}(t)=\rho_e^{01}(0) e^{-\sum_{kl}W_{kl}(t)}.
    \label{Eq:General_MB_solution}
\end{equation}
Each nuclear spin pair $k,l$ contributes to the echo decay with an amplitude,
\begin{equation}
    \alpha_{kl}^2 = \bigg(\frac{2\Delta_{kl}b_{kl}}{\Delta_{kl}^2+b_{kl}^2}\bigg)^2,
    \label{Eq:alphasq}
\end{equation}
and frequency,
\begin{equation}
    f_{kl} = \frac{1}{4}\sqrt{\Delta_{kl}^2+b_{kl}^2}.
    \label{Eq:freq}
\end{equation}
We find that the amplitude, $\alpha_{kl}^2$, reaches a maximum of 1.0 when $|b_{kl}|=|\Delta_{kl}|$, and vanishes when either $b_{kl}$ or $|\Delta_{kl}|=0$ or in the limit that $|\Delta_{kl}|\rightarrow \infty$, consistent with previous results.~\cite{Jeschke:2023,Suchaneck:2025} \textcolor{black}{To demonstrate a benefit of a TCL treatment, we derive the fourth-order correction to the TCL equation to assess the accuracy. Details on the derivation of the TCL4 equation are shown in the SI. The TCL4 equation is,
\begin{equation}
    \rho_e^{01}(t) = \frac{1}{2}\textrm{exp}\bigg[\sum_{kl} -4\bigg(\frac{b_{kl}\Delta_{kl}}{b_{kl}^2+\Delta_{kl}^2}\bigg)^2 \textrm{sin}^4\bigg(\frac{t}{4}\sqrt{\Delta_{kl}^2+b_{kl}^2}\bigg)-12\bigg(\frac{b_{kl}\Delta_{kl}}{b_{kl}^2+\Delta_{kl}^2}\bigg)^4 \textrm{sin}^8\bigg(\frac{t}{4}\sqrt{\Delta_{kl}^2+b_{kl}^2}\bigg)\bigg],
    \label{eq:TCL4}
\end{equation}
where all of the parameters have been defined previously.}

\begin{figure}[ht!]
    \centering
    \includegraphics[width=1.0\linewidth]{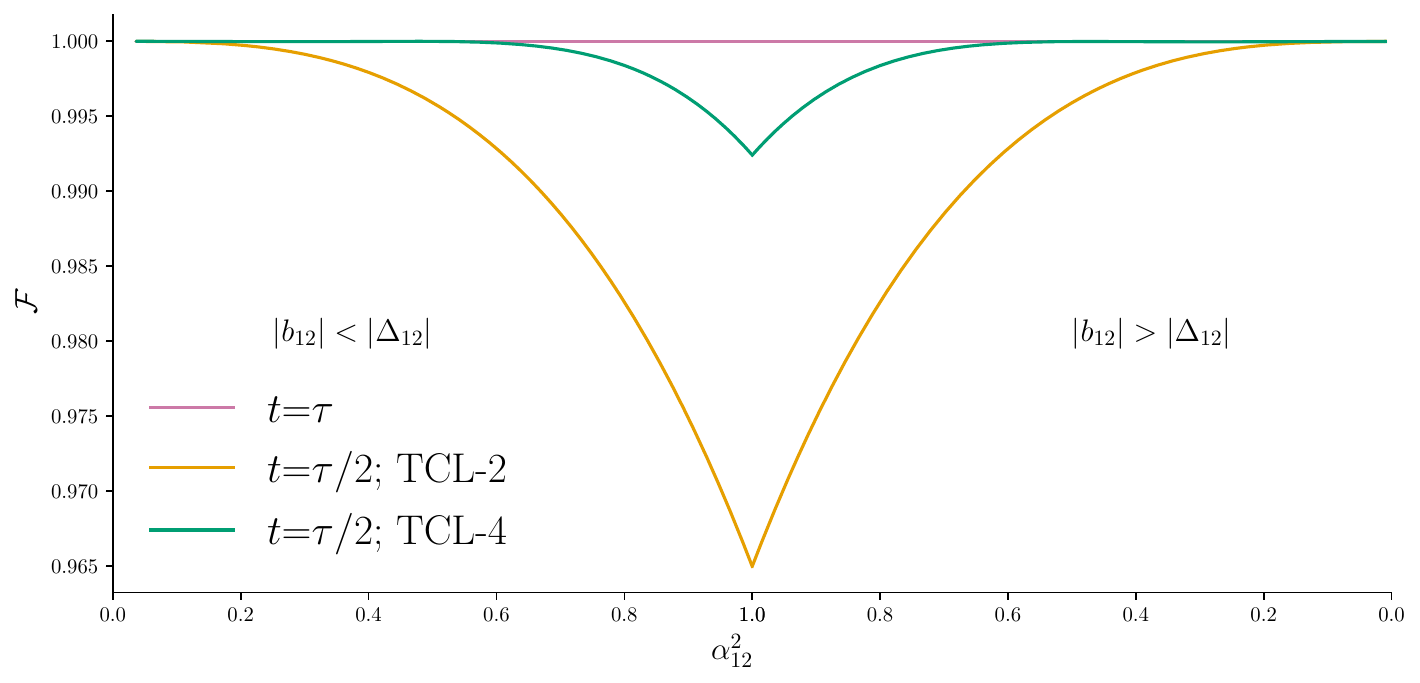}
    \caption{Fidelity between the TCL2 \textcolor{black}{and TCL4} and numerically exact density matrices as a function of $\alpha_{12}^2$. The left side of the figure represents when $|b_{12}|<|\Delta_{12}|$ and the right side when $|b_{12}|>|\Delta_{12}|$.}
    \label{fig:TCL-analysis}
\end{figure}

We compare the dynamics of our TCL2 \textcolor{black}{and TCL4} equation\textcolor{black}{s} to numerically exact simulations of one electron interacting with two nuclear spins by computing the fidelity between the two density matrices,
\begin{equation}
    \mathcal{F}(\hat{\rho},\hat{\sigma}) = \bigg(\textrm{Tr}[\sqrt{\sqrt{\hat{\rho}}\hat{\sigma}\sqrt{\hat{\rho}}}]\bigg)^2.
    \label{Eq:Fidelity}
\end{equation}
The coherence of the electron spin periodically returns to a maximum value of 1 after a time interval $\tau=\frac{1}{f_{12}}$, and reaches a minimum value at $\tau/2$. Figure~\ref{fig:TCL-analysis} shows the fidelity computed at $t=\tau/2$ and $t=\tau$ to evaluate the error in modulation depth and frequency, respectively. Details on the numerically exact simulations and explicit echo profiles are shown in the SI.
The fidelity at $\tau$ remains at 1 for all values of $\alpha_{12}^2$, indicating that our TCL2 \textcolor{black}{and TCL4} equation\textcolor{black}{s} correctly predict the frequency, $f_{12}$, of the coherence. The fidelity at $\tau/2$ decreases as $\alpha_{12}^2$ increases, approaching a minimum of \textcolor{black}{0.965 and 0.992} when $\alpha_{12}^2$ reaches a maximum of 1.0 \textcolor{black}{for TCL2 and TCL4, respectively}. Despite the deviations from the numerically exact value of the coherence \textcolor{black}{as $\alpha_{12}^2$ approaches it's maximum}, \textcolor{black}{the fidelity of the TCL2 and TCL4 density matrices remain high} making \textcolor{black}{the TCL approach} a good \textcolor{black}{model} of the contribution a nuclear-spin pair makes to electron dephasing in the Hahn-echo experiment. 

\begin{figure}[ht!]
    \centering
    \includegraphics[width=1.0\linewidth]{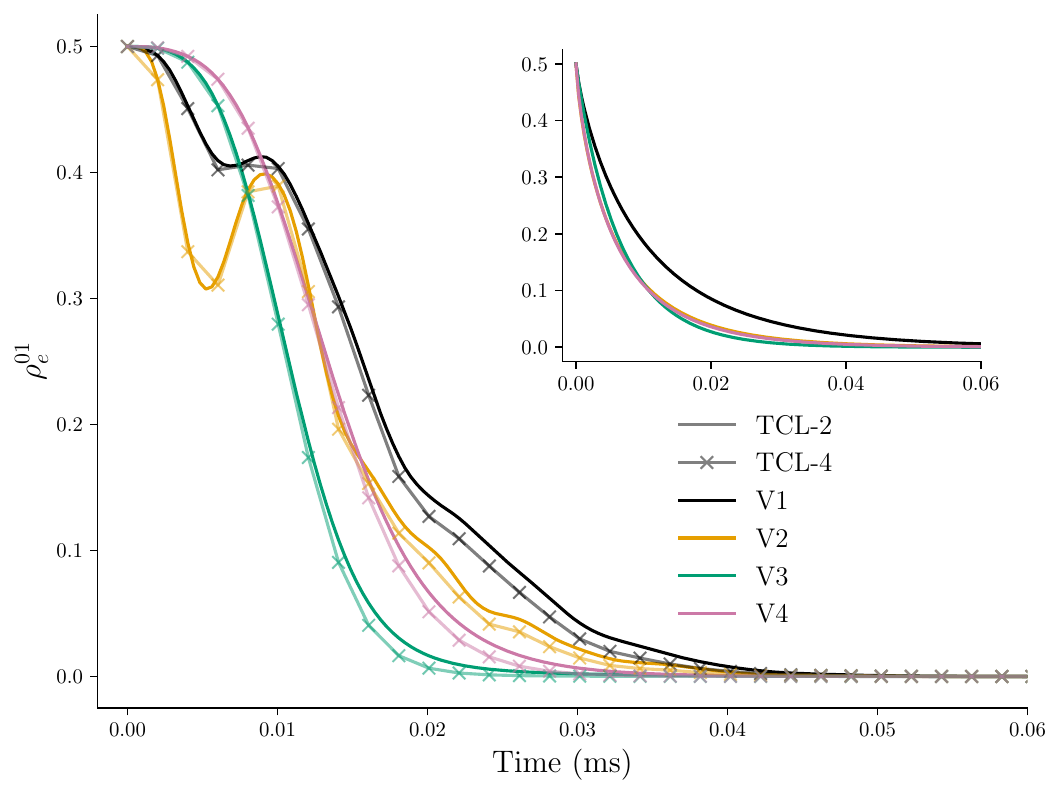}
    \caption{Echo decay of the vanadium-oxo molecules interacting with a bath of hydrogen spins\textcolor{black}{, computed using TCL2 and TCL4}. \textcolor{black}{Experimental data is reproduced from the fit parameters in Ref.~\citenum{Graham:2017} and shown in the inset.}}
    \label{fig:V_other_molecules}
\end{figure}

We now demonstrate the factorized TCL2 \textcolor{black}{and TCL4} equation\textcolor{black}{s} in Eq\textcolor{black}{s}.~\ref{Eq:General_MB_solution} \textcolor{black}{and~\ref{eq:TCL4}} by analyzing a series of previously studied vanadium-oxo molecules,  [VO(C$_3$H$_6$S$_2$)$_2$]$^{2-}$, [VO(C$_5$H$_6$S$_4$)$_2$]$^{2-}$, [VO(C$_7$H$_6$S$_6$)$_2$]$^{2-}$ and [VO(C$_9$H$_6$S$_8$)$_2$]$^{2-}$, labeled as \textbf{V1}, \textbf{V2}, \textbf{V3} and \textbf{V4}, respectively.~\cite{Graham:2017} The full molecular structures are shown in the SI. From \textbf{V1} to \textbf{V4} the spin active hydrogens become progressively further from the electron spin located on the vanadium metal center. We compute the hyperfine couplings between the unpaired electron on the vanadium ion and the distal hydrogen spins using density functional theory (DFT), and compute the dynamics of the electron spin coherence using Eq\textcolor{black}{s}.~\ref{Eq:General_MB_solution} \textcolor{black}{and~\ref{eq:TCL4}}.
\begin{figure}[ht!]
    \centering
    \includegraphics[width=1.0\linewidth]{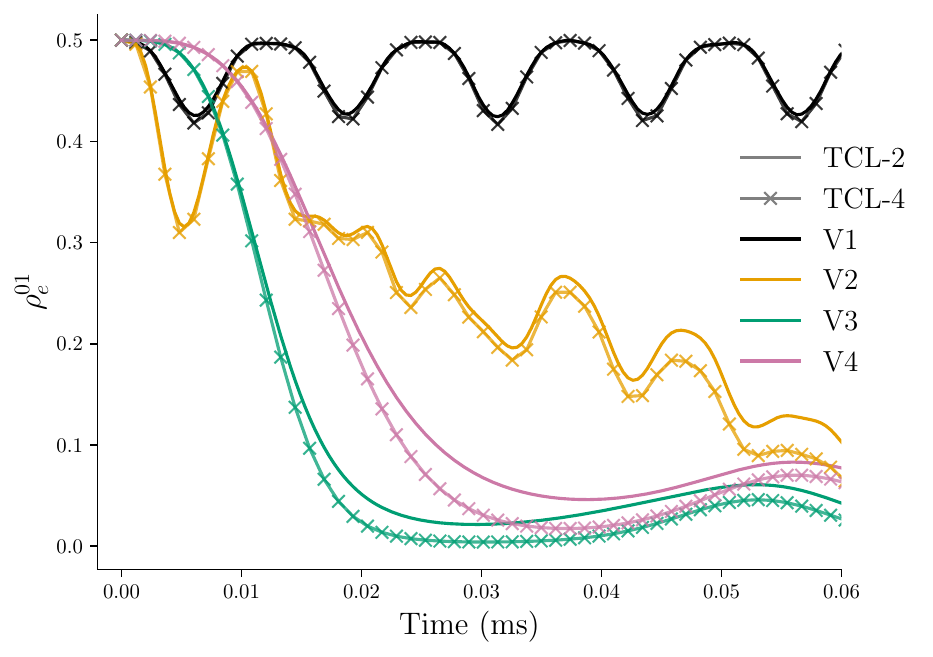}
    \caption{Echo decay for four different vanadium-oxo molecules using the factorized TCL2 \textcolor{black}{and TCL4} approaches.}
    \label{fig:Vseries}
\end{figure}
The dynamics of the electronic state in these complexes are shown in Figure~\ref{fig:Vseries}. Here we do not see complete decoherence but rather so-called residual coherence, which is expected for small nuclear spin baths as demonstrated in previous computational studies.~\cite{Chen:2020} Experimental data taken on these molecules shows that coherence decays to zero within the time scale shown in Fig.~\ref{fig:Vseries}.~\cite{Graham:2017} To better represent the experiment we embed the molecule in a bath of hydrogen spins, simulating hydrogens present on nearby vanadium-oxo\textcolor{black}{, counter-ion,} or solvent molecules. We add those contributions to Eq.~\ref{Eq:General_MB_solution}, and show these dynamics in Figure~\ref{fig:V_other_molecules}.
In the presence of additional hydrogen spins, we see that the echo decays completely, with \textbf{V1} taking the longest to decay to zero, consistent with experimental data.~\cite{Graham:2017} \textcolor{black}{At early times in the echo profile, the experimental dephasing is highly oscillatory, likely due to pseudosecular interactions between nuclear spins that we do not consider in our theory.~\cite{Sousa:2009} For clarity we have only included the stretched exponential decay of the experimental coherence. The factorization of nuclear spins into pairs introduces error at longer times; however, the pair approximation is still capable of qualitatively capturing experimental trends, likely because the important dynamics of dephasing for the molecular data occur in the timeframe where the pair approximation holds. Additionally, the TCL2 and TCL4 equations could be solved numerically for clusters of greater than two nuclear spins to improve accuracy in regimes where the pair approximation fails.}

Notably in the above data, we did not consider decoherence contributions from heteronuclear spin pairs. Molecular qubit candidates often include transition metal ions or deuterated solvent molecules with $I>\frac{1}{2}$ nuclei, which can interact with hydrogen spins via dipolar interactions. We can use the TCL2 framework to evaluate the impact of heteronuclear spin pair fluctuations on the echo decay. 
\begin{table*}[ht!]
    \centering
    \begin{tabular}{@{\extracolsep{4pt}}l c c}
        \hline\hline
         $I$ & Example Nuclei & $W(t)$ \\
        \hline 
        $1$  & ${^2}$D & 10$^{-14}$\\
        $3/2$ & ${^{63}}$Cu & 10$^{-11}$ \\
        $5/2$  & ${^{55}}$Mn & 10$^{-12}$ \\
        $7/2$  & ${^{51}}$V & 10$^{-12}$ \\  
        \hline\hline
    \end{tabular}
    \caption{Contributions to the echo decay for spin pairs with one spin $I>\frac{1}{2}$.}
    \label{tab:hnuc_pairs}
\end{table*}
We solve Eq.~\ref{Eq:Wtintegral} for several possible nuclear spins with different $I$ interacting with a spin-$\frac{1}{2}$ nucleus. For different spin states, we compute the heteronuclear correlation function and solve Eq.~\ref{Eq:Wtintegral} to obtain the heteronuclear pair contribution to spin decoherence. Table~\ref{tab:hnuc_pairs} shows the order of magnitude of $W(t)$, which can be inserted into Eq.~\ref{Eq:General_MB_solution} to simulate the damping of the dynamics from a heteronuclear spin pair. For the heteronuclear spin pairs considered, $W(t)$ is vanishingly small, indicating that spin-spin interactions between heteronuclear spin pairs, such as vanadium and hydrogen, do not contribute to the echo decay for spin-spin electron dephasing within the secular spin Hamiltonian at high magnetic field. \textcolor{black}{This result is consistent with previous studies of heteronuclear spin flip-flops.~\cite{Chekhovich:2017,Cywinski:2009,Liu:2007}}

We have presented a non-Markovian ME approach to pure dephasing in molecular spin systems at low temperature in a large magnetic field \textcolor{black}{where pure dephasing dominates}. \textcolor{black}{Under these conditions, the secular spin Hamiltonian is a good model for the dynamics; however, incorporating decoherence beyond the pure dephasing limit will require careful treatment of the interaction picture representation of the electron spin beyond the secular approximation.} We analyze the performance of the TCL2 equation against numerically exact simulations for a three-spin system and we find good agreement. \textcolor{black}{We increase the accuracy of the TCL approach by including the fourth order correction.} Specifically, the frequency of oscillations generated by our TCL2 approach agrees with the numerically exact solution for all parameter regimes tested. The TCL2 \textcolor{black}{and TCL4} approach\textcolor{black}{es} underestimate the modulation depth of the electron coherence \textcolor{black}{at larger values of $\alpha_{12}^2$}, but provide a good prediction of the contribution of each nuclear spin pair to the decoherence. We then extend the TCL2 \textcolor{black}{and TCL4} technique to molecular spin systems using a factorization approach of the spin-spin correlation functions, computing hyperfine coupling tensors with electronic structure calculations. Here we see trends in the molecular series which agree with previous computational studies, where the smaller molecule retains more coherence than larger molecules. To agree with previous experimental studies, we embed the molecules in a bath of hydrogen spins and simulate complete decoherence of the electronic spin state within the experimentally observed timescale\textcolor{black}{, highlighting the accuracy of the factorization approach in the short-time window where the pair approximation holds}. In these molecular studies, we neglect any electron decoherence contributions from heteronuclear spin-spin interactions. To justify this approximation, we compute $W(t)$ for several nuclear spins with $I>\frac{1}{2}$ commonly present in molecular qubits, and show that these contributions are vanishingly small. 

The promise of molecular QIS relies on controlling and leveraging the diversity of molecular conformations and electronic structures. Relating the \textit{ab initio} electronic properties to the time-dependent behavior of molecules in magnetic fields at low temperatures is essential for robust understanding of electron-nuclear spin decoherence in molecules. Our results directly relate molecular properties, including hyperfine tensors and nuclear dipolar coupling, to low-temperature decoherence dynamics in molecular spin systems. Importantly, the approach requires only a single electronic structure calculation per species. This is important in cases where the molecule cannot be described by DFT, but requires more costly techniques to simulate the correlated electronic structure. Furthermore, our technique will allow for greater understanding of how electron delocalization and correlation amplify or diminish nuclear-spin decoherence effects. \textcolor{black}{Furthermore, by using an OQS framework to describe the Hahn-echo decay of electron coherence, we can begin to incorporate thermal effects that go beyond the pure dephasing limit to describe the temperature dependence of $T_2$.}

\section{Computational Details}

We obtain hyperfine tensors for \textbf{V1} - \textbf{V4} from electronic structure calculations using the ORCA 6 quantum chemistry package.~\cite{Neese:2020} Following previous work, we optimize the geometries using density functional theory with B3LYP and density fitting, with a def2-triple-$\zeta$ polarized basis for hydrogen, sulfur, carbon, and oxygen and a core-polarized CP(PPP) basis for vanadium.~\cite{Chen:2020,Chmela:2018,Hellweg:2007,Weigend:2005,Weigend:2006,Parr:1988a,Becke:1988a} The vanadium-oxygen bond is oriented to lie along the $z$-axis, and we compute hyperfine tensors from the optimized geometries and use the $zz$ component in Eq\textcolor{black}{s}.~\ref{Eq:General_MB_solution} \textcolor{black}{and~\ref{eq:TCL4}} for each hydrogen spin. We include contributions to the hyperfine tensors from the dipolar coupling between the electron and nuclear spins along with the isotropic Fermi contact term. We include a bath of random hydrogen spins for comparison to experimental results. \textcolor{black}{To generate a physically reasonable bath, we place the molecule at the center of a cube with a 40 \AA~edge length. We then estimate the concentration of solvent molecules in the cube using the density of a common solvent molecule, N,N-dimethylformamide-d7 (DMF) with an isotopic purity of 99\%, or 1\% undeuterated hydrogens. We increase the concentration of the undeuterated hydrogens to account for the presence of undeuterated counter-ion molecules, tetraphenylphosphonium (Ph$_4$P$^+$) present in the experimental system. The interactions between the solvent hydrogen spins and electron spin are computed using the point-dipole approximation and added to Eqs.~\ref{Eq:General_MB_solution} and~\ref{eq:TCL4}.} The randomized coordinates of the extra hydrogen spins were generated once and used for each of \textbf{V1} - \textbf{V4}.

\section{Associated Content}

\subsection{Data Availability Statement}

The data and codes underlying this study are openly available at \url{https://doi.org/10.5281/zenodo.18246098}.

\subsection{Supporting Information}
Details on the derivation of the TCL\textcolor{black}{2 and TCL4} equation\textcolor{black}{s}, the spin Hamiltonian, the error analysis, \textcolor{black}{ the factorization of spin pairs, comparison to the APPA,} and molecular systems. 

\section{Acknowledgements}

KHM acknowledges start-up funding from the University of Minnesota. 

\section{References}

\bibliography{main}

\end{document}